\documentstyle[prl,aps,eqsecnum,epsfig,multicol]{revtex}
\begin{document}
\draft

%%%%%%%%%%%%%%%%%%%%%%%%%%%%%%%%%%%%%%%%%%%%%%%%%%%%%%%%%%%%%%%
% Title
%

\title{Measurement of single electrons and 
implications for charm production \\
in Au+Au collisions at $\sqrt{s_{_{NN}}}$~=~130~GeV }

% NOTE:  Brant will add the author list later.  Its length
% does not count toward the length estimate.  Instead, 24 
% lines are counted toward the length to approximate a 
% typical set of several authors and a few institutions.
%
% Use the following 5 lines exactly as written:

%% Author and Institution List for PPG003
\author{
K.~Adcox,$^{40}$
S.{\,}S.~Adler,$^{3}$
N.{\,}N.~Ajitanand,$^{27}$
Y.~Akiba,$^{14}$
J.~Alexander,$^{27}$
L.~Aphecetche,$^{34}$
Y.~Arai,$^{14}$
S.{\,}
% ...
\\(PHENIX Collaboration)\\
}
\address{
% ...
$^{3}$Brookhaven National Laboratory, Upton, NY 11973-5000, USA\\
% ...
$^{14}$KEK, High Energy Accelerator Research Organization, Tsukuba-shi, Ibaraki-ken 305-0801, Japan\\
% ...
$^{27}$Chemistry Department, State University of New York - Stony Brook, Stony Brook, NY 11794, USA\\
% ...
$^{34}$SUBATECH (Ecole des Mines de Nantes, IN2P3/CNRS, Universite de Nantes) BP 20722 - 44307, Nantes-cedex 3, France\\
% ...
$^{40}$Vanderbilt University, Nashville, TN 37235, USA\\
% ...
}

\date{\today}        
\maketitle

%%%%%%%%%%%%%%%%%%%%%%%%%%%%%%%%%%%%%%%%%%%%%%%%%%%%%%%%%%%%%%%
% Abstract
%

\begin{abstract}

Transverse momentum spectra of electrons from Au+Au collisions
at $\sqrt{s_{_{NN}}} = 130$~GeV have been measured at midrapidity
by the PHENIX experiment at RHIC.
The spectra % from central and minimum bias collisions
show an excess above the background from photon conversions and light hadron
decays. % for $p_T > 0.6$~GeV/c
The electron signal is consistent with that expected
from semi-leptonic decays of charm.
The yield of the electron signal $dN_e/dy$ for $p_T > 0.8$~GeV/c 
is $0.025\pm0.004$(stat.)$\pm0.010$(sys.) in central collisions, and
the corresponding charm cross section is $380\pm60$(stat.)$\pm200$(sys.)~$\mu$b
per binary nucleon-nucleon collision.

\end{abstract}

\pacs{PACS numbers: 25.75.Dw}

\begin{multicols}{2}   % This is needed only for the "multicols" style
\narrowtext            % This is needed only for the "multicols" style
%NOTES: 
%1.  For PRL do not use section headings.
%
%2.  Do not worry about indenting the first line of a paragraph.  Just
%    insert a blank line between paragraphs.  Similarly, if you want 
%    an equation to stay within a paragraph, do not put a blank line
%    before or after the equation.
%
%3.  Do not embed figures or tables; place them all at the end (see below).
%
%4.  Name all references and use "\cite{refname}" in the text to cite them.
%    (The RevTeX macro will replace this with "[1]" in proper PR style.)
%
%5.  The list of references must be ordered in the same sequence as they
%    occur in the text.
%
%6.  Use our standard aknowldegement below as the last paragraph of your
%    text.  (Yes, it does count toward the length!).
%
%%%%%%%%%%%%%%%%%%%%%%%%%%%%%%%%%%%%%%%%%%%%%%%%%%%%%%%%%%%%%
% general introduction
%
% \marginpar{{\small \em Intro}}
%
%\twocolumn
%
% P1 Introduction
%
In this letter, we report the first measurement of
single electron spectra, $(e^{+}+e^{-})/2$, in Au+Au collisions at
$\sqrt{s_{_{NN}}}=130$~GeV % by the PHENIX experiment
at the Relativistic Heavy Ion Collider (RHIC).
The measurement of single leptons at high
transverse momentum ($p_T \gtrsim 1$~GeV/c) is a useful way to
study heavy-quark production, % which is
an important probe of
hot and dense matter created in high energy heavy ion collisions.
Charm production is sensitive to the
initial state gluon density\cite{Appel,XNWang}. Nuclear and medium effects,
such as shadowing and charm quark energy loss\cite{Lin,Kharzeev},
can be studied
by comparison of charm production
in $AA$, $pA$, and $pp$ collisions.
Measurement of charm is important for understanding
J/$\psi$ suppression %, which is
(a proposed signal of the deconfinement
phase transition\cite{Masui,NA50})
and the dilepton mass distribution in $ 1 < M_{l^+l^-} < 3 $~GeV,
where lepton pairs from charm make significant contributions\cite{Vogt}.
In $pp$ collisions at the ISR ($\sqrt{s}=30 - 63$~GeV),
production of single electrons was observed
% at a signal level of
($e/\pi \sim 10^{-4}$)
for $p_T > $
1~GeV/c\cite{CCRS74,CCRS76,Perez,Basile}, and % has been 
interpreted as evidence of open charm
production\cite{Hinchliffe_Smith}.
In $pp$ collisions at RHIC energies,
the signal level is expected to be higher, since
charm production increases with $\sqrt{s_{_{NN}}}$ faster than
pion production. We recently observed suppression of 
high $p_T$ pion production in Au+Au collisions at RHIC
relative to binary nucleon-nucleon ($NN$) collision
scaling\cite{PPG003}.
If charm production scales with $NN$ collisions,
as expected in the absence of nuclear effects,
the $e/\pi$ ratio will be even higher in
Au+Au collisions at RHIC.

%%%%%%%%%%%%%%%%%%%%%%%%%%%%%%%%%%%%%%%%%%%%%%%%%%%%%%%%%%%%%%%%%%%%%%%%%
% 
% P2 Data set and detector
%
Data used for this analysis were recorded by the PHENIX
west-arm spectrometer\cite{hamagakiQM01}
($\Delta \phi = 90^{o}$ in azimuth,
$|\eta|<0.35$ in pseudo-rapidity), which
consisted of a drift chamber (DC), a layer of
pad chambers (PC1),
a ring imaging Cerenkov detector (RICH),
and a lead-scintillator
electromagnetic calorimeter (EMCAL).
The trigger was provided by % pairs of
beam-beam counters (BBC) and
zero-degree calorimeters (ZDC).
ZDC and BBC signals were combined to select
centrality: 
central (0-10\%), peripheral (60-80\%), and
minimum bias (0-92\%)\cite{PPG001}.

%
% P3 Event and electron track selection
%
The analysis uses 1.23 M minimum bias events with
vertex position $|z| < 30$ cm.
Charged particle tracks are reconstructed by the DC
and the PC1 %, and their momenta $p$ are determined by the DC with a
with a momentum resolution % of 
$\delta p/p \simeq 0.6\% \oplus 3.6\%\ p$~(GeV/c).
% The absolute momentum scale is known to better than 2\%.
Tracks are confirmed by a matching hit in the EMCAL,
which measures the energy $E$ deposited with a resolution of
$8.2\%/\sqrt{E({\rm GeV})} \oplus 1.9\%$ for test beam electrons.
Electron identification is performed using the RICH and
the EMCAL\cite{hamagakiQM01}.
The RICH is filled with 1 atm CO$_2$ % radiator 
and detects
on average 10.8 photo-electrons per electron track, while a pion with
$p<4.7$~GeV/c produces no signal.
It is required that at least 3 % RICH PMT hits
RICH hits
are associated with
the track and that their hit pattern is consistent with that of an
electron track. After these cuts,
a clear electron signal is observed
as a narrow peak centered at $E/p$ = 1.0.
% 1.0 in the ratio of $E/p$.
%, where
% $E$ is the energy measured by the EMCAL and $p$ is the momentum
% measured by the DC.
%
% The following explanation of E/p cut may be removed if
% space is too short.
%
We select tracks in the peak as electron candidates.
The $E/p$ cut reduces hadron background, and removes conversion
electrons created far from the vertex.
A hadron deposits
only a fraction of its energy in the EMCAL, and
the momentum of an off-vertex conversion electron
is reconstructed incorrectly.
The remaining background, about 10\% of the electron candidates, is caused
by accidental association of RICH hits with hadron tracks.
The background level is measured
statistically by an event mixing method,
and is subtracted from the yield.

%
% P4 Acceptance and corrections
%
The electron acceptance ($\sim 7.4$ \% of $dN/dy$) and
efficiency ($\sim 60$ \%) are determined using
a detailed GEANT\cite{GEANT} simulation, which satisfactorily
reproduces the detector response.
Additionally, a multiplicity dependent efficiency loss
due to detector occupancy is evaluated by embedding
simulated electrons into real events. This efficiency loss is
$27 \pm 4$\% ($4 \pm 2$\%) for central (peripheral) collisions and 
has no significant $p_T$ dependence.

\begin{figure}
%\begin{center}
\centerline{\epsfig{file=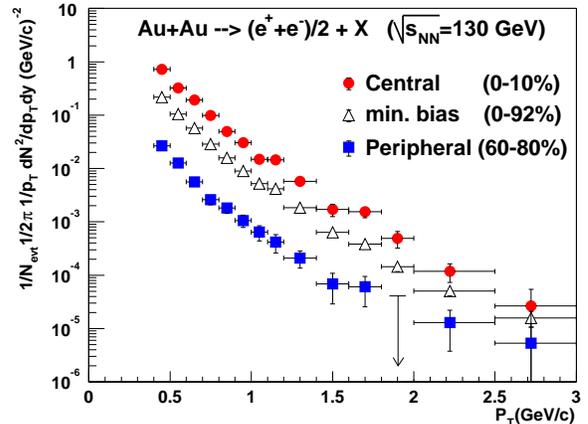,width=1.0\linewidth}}
\caption[]{
	Transverse momentum spectra of electrons
	in PHENIX from Au+Au collisions at $\sqrt{s_{_{NN}}}$=130~GeV.
	% The centrality bins are indicated in the figure.
	}
%\end{center}
\label{fig:e_spectra} 
\end{figure}

%%%%%%%%%%%%%%%%%%%%%%%%%%%%%%%%%%%%%%%%%%%%%%%%%%%%%%%%%%%%%%%%%%%%%%%%
% P5 describe Fig.1
%
Fig.~\ref{fig:e_spectra} shows the $p_T$ distributions of electrons
in PHENIX for central, minimum bias, and peripheral collisions.
Errors in the figure are statistical.
The overall systematic uncertainty, which is the
quadratic sum of several few percent effects, is about 11\%.
Expected sources of electrons are
(1) Dalitz decays of $\pi^0$, $\eta$, $\eta'$, $\omega$, and $\phi$,
(2) di-electron decays of $\rho$, $\omega$, and $\phi$,
(3) photon conversions,
(4) kaon decays ($K^{0,\pm} \rightarrow \pi e \nu$),
(5) semi-leptonic decay of charm, and 
(6) other contributions such as bottom decays and thermal di-leptons.
In this analysis, sources (1)-(4) are considered to be background.

%%%%%%%%%%%%%%%%%%%%%%%%%%%%%%%%%%%%%%%%%%%%%%%%%%%%%%%%%%%%%%%%%%%%%%%%
% the background cocktail
%
% P6 Dalitz background
%
We have calculated the contributions from Dalitz and di-electron decays
with a hadron decay generator.
PHENIX has measured the $p_T$ distributions of $\pi^{\pm}$ in
$0.2 < p_T < 2.2$~GeV/c\cite{PPG006} and of $\pi^{0}$ in
$ 1< p_T < 4$~GeV/c\cite{PPG003}.
Since the $\pi^{\pm}$ and $\pi^0$ data are consistent in the
overlapping region, we fit a power law function to the combined data
sets to determine the input $\pi^{0}$ spectrum for the decay generator.
The $p_T$ distribution of any other hadron $h$ is obtained from the
$\pi^{0}$ spectrum by replacing $p_T$ with
$\sqrt{p_T^2 + m_{h}^{2} - m_{\pi^0}^2}$.
The shapes of the resulting $p_T$ spectra of $K^{\pm}$, $p$, and 
$\bar{p}$ agree
with the PHENIX measurements\cite{PPG006} within 20\%.
% By construction
In this parameterization
$h/\pi^0$ ratios approach constants at
high $p_T$.
We assume the following asymptotic ratios to fix the relative
normalizations:
$\eta/\pi^0 = 0.55$, $\eta'/\pi^0 = 0.25$,
$\rho/\pi^0 = \omega/\pi^0 = 1.0$, $\phi/\pi^0 = 0.40$.
Except for the $\phi$, these ratios are taken from proton beam
data of SPS, FNAL, and ISR experiments\cite{eta_ratio,omega_ratio}.
The $\eta/\pi^0$ ratio is consistent with % that measured in Pb+Pb
a measurement in Pb+Pb
collisions at SPS\cite{WA98}.
The $\phi/\pi^0$ ratio is based on % a recent STAR measurement of the
the integrated ratio $\phi/h^{-} \sim 0.02$ in Au+Au collisions
at $\sqrt{s_{_{NN}}}$ = 130 GeV\cite{STAR_phi}.
We assign to each ratio a conservative
systematic uncertainty of 50\%.

%%%%%%%%%%%%%%%%%%%%%%%%%%%%%%%%%%%%%%%%%%%%%%%%%%%%%%%%%%%%%%%%%%%%%%%%%%
%
% P7 photon background
%
Photon conversions are evaluated using a combination
of the GEANT simulation and the hadron decay generator.
% Since Dalitz decays are internal conversions of real photon decays,
% the $p_T$ spectra of externally converted electrons are almost
% identical to those from the Dalitz decays.
% Therefore, a good approximation of the conversion spectra
% can be made by scaling the Dalitz decay spectra by an appropriate factor.
%
% (The above two sentences in release 2 is changed the sentence below
% to save  a few extra lines)
%
% The following part is slightly changed to incorporate comment B11.
% The new text is based on Ralf's suggestion.
%
Since $p_T$ spectra of externally converted electrons are
similar to those from Dalitz decay, 
the conversion spectra can be approximated by
scaling the Dalitz decay spectra by an experiment specific factor,
$R_{conv}$ = Conversion/Dalitz. $R_{conv}$ 
is evaluated using the GEANT simulation
and is cross-checked by comparing the relative
yield of reconstructed Dalitz and conversion pairs in
the simulation and in data.
The simulation shows that $R_{conv}$ has only a weak
$p_T$ dependence, primarily due to the energy dependence of the pair
creation cross section.
$R_{conv}$ is parameterized as
$(1.9 \pm 0.2)\times (1 - 0.0718 \times p_T^{-0.76})$.

%
% P8 Kaon decay contributions (Vince wants this to be a paragraph)
%
Background from kaon decays is also  evaluated using the GEANT simulation,
and is found to be negligible.

\begin{figure}
%\begin{center}
\centerline{\epsfig{file=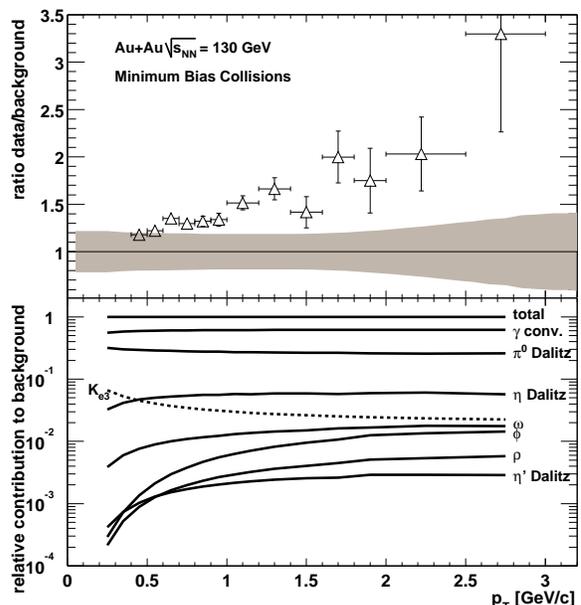,width=1.0\linewidth}}
\caption[]{
	Ratio of the electron
	data to the calculated background as a function of $p_T$
	in minimum bias collisions (upper panel) and
	relative contributions to the background from 
	various sources (lower panel).
	The curves for $\omega$ and $\phi$ show the sum of the
	Dalitz and the di-electron decay modes.
	}
%\end{center}
\label{fig:ratio} 
\end{figure}

%%%%%%%%%%%%%%%%%%%%%%%%%%%%%%%%%%%%%%%%%%%%%%%%%%%%%%%%%%%%%%%%%%%%%%%%%%
%
% P9 Describe Fig.2 upper panel
%
The upper panel of Fig.~\ref{fig:ratio}
shows the ratio of the measured electrons to
the calculated background versus $p_T$ for minimum bias events.
The shaded region is the quadratic sum of
systematic errors in the electron measurement and
in the background. The latter includes uncertainties in the normalization
and the shape of the $\pi^0$ spectrum, in the
$h/\pi^0$ ratios, and in $R_{conv}$.
A significant electron excess above the background
is observed for $p_T>0.6$~GeV/c. Central collisions
show a similar excess. % for $p_T>0.6$~GeV/c.
% Peripheral collisions do not have sufficient statistics to
% see a signal in this analysis.
The peripheral collision data sample lacks sufficient statistics to
reveal a signal in this analysis.

%%%%%%%%%%%%%%%%%%%%%%%%%%%%%%%%%%%%%%%%%%%%%%%%%%%%%%%%%%%%%%%%%%%%%%%%
%
% P10 Discussion on uncertainties in the cocktail defended
%
Fractional contributions to the background
are shown in the lower panel of Fig.~\ref{fig:ratio}.
More than 80\% of the background is from $\pi^0$ decay,
directly from the Dalitz decay or indirectly from photon conversion.
The $\pi^0$ spectrum is well constrained by the PHENIX measurement.
The next most important background source is $\eta$ decay.
% The 50\% uncertainty assigned to the ratio $\eta/\pi^0 = 0.55$
% is conservative, since an even larger $\eta/\pi^0$ ratio would
% imply that 
% $\eta$ production at high $p_T$ is close to or larger than
% primary $\pi^0$ production.
%
% (after the referee's comment A10, Vince (+Ralf +Bill) changes this sentence
% to the one below.
%
Given the assigned systematic error, the upper limit of the high
$p_T$ asymptotic $\eta/\pi^0$ ratio is 0.83. Since this ratio,
corrected for feed-down, would imply that the primary
$\eta/\pi^0 \sim 1$, this provides a conservative limit on contributions
from $\eta$'s.
Contributions from all other hadrons combined are only a few
percent of the total.

\begin{figure}
%\begin{center}
\centerline{\epsfig{file=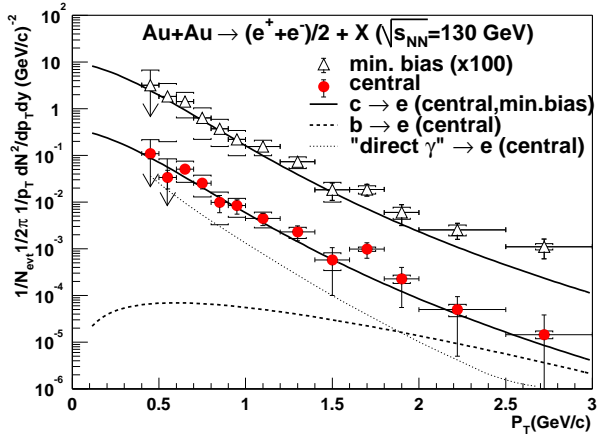,width=1.0\linewidth}}
\caption[]{
	The background-subtracted electron spectra for
	minimum bias (0-92\%) % (scaled up by factor 100)
	and central (0-10\%) collisions compared with the
	expected contributions from open charm decays.
	Also shown, for central collisions only, are the
	expected contribution from bottom decays (dashed) and
	the conversion electron spectrum
	from a direct photon prediction (dotted).
	}
%\end{center}
\label{fig:signal_spectrum} 
\end{figure}

%%%%%%%%%%%%%%%%%%%%%%%%%%%%%%%%%%%%%%%%%%%%%%%%%%%%%%%%%%%%%%%%%%%%%%%%
%
% P11 Background subtracted spectrum
%
Background-subtracted electron spectra are shown in
Fig.~\ref{fig:signal_spectrum}.
The error bars on the data points represent the
statistical errors, while the systematic error due to the
background subtraction is indicated by brackets. % or arrows.
The integrated yield of the electron signal
$dN_e/dy$ for $p_T>0.8$~GeV/c is $0.025\pm0.004$(stat.)$\pm0.010$(sys.) for
central collisions and is $0.0079\pm0.0006$(stat.)$\pm0.0034$(sys.) for
minimum bias collisions.

%%%%%%%%%%%%%%%%%%%%%%%%%%%%%%%%%%%%%%%%%%%%%%%%%%%%%%%%%%%%%%%%%%%%%%%%
% P12
% PYTHIA charm model
%
Semi-leptonic decay of charmed hadrons is an expected source of
the electron signal.
We use the event generator PYTHIA\cite{PYTHIA} to
estimate electron spectra from charm decay.
We tuned the parameters\cite{PYTHIA_par} of PYTHIA such that charm 
production data at SPS and FNAL\cite{charm_data}
and single electron data at the ISR\cite{CCRS76,Perez,Basile}
are well reproduced.
The charm production cross section in $pp$ collisions
from this PYTHIA calculation
is $\sigma_{c\bar{c}}=330\mu$b at $\sqrt{s}=$130~GeV.
%
% Compare the data with PYTHIA charm calculation
%
The electron spectrum in Au+Au collisions is then calculated as
$EdN_e/dp^3 =T_{\small AA} \times Ed\sigma_{e}/dp^3$, where
$Ed\sigma_e/dp^3$ is the electron spectrum from charm decay
calculated with PYTHIA, and
$T_{\small AA}$ (listed in Table~\ref{tab:yield})
is the nuclear overlap integral calculated from a
Glauber model\cite{PPG003}.
The calculated electron spectra shown in
Fig.~\ref{fig:signal_spectrum} are in reasonable agreement with the data.

%%%%%%%%%%%%%%%%%%%%%%%%%%%%%%%%%%%%%%%%%%%%%%%%%%%%%%%%%%%%%%%%
% P13 Other contributions, which we will ignore
%
Before attributing the entire electron signal to open charm decays,
it is necessary to quantify contributions from other possible sources.
%
% bottom, J/PSI, and Drell-Yan contribution.
%
An analogous
PYTHIA estimate of the bottom decay
contribution % for central collisions
is shown in Fig.~\ref{fig:signal_spectrum}. % as a thick solid curve.
It becomes significant only above the measured $p_T$ range.
Expected contributions from $J/\Psi$ 
and Drell-Yan are negligible.
%
% Thermal radiation contributions
%
In Pb+Pb collisions at SPS, direct photons\cite{WA98}
and an enhanced yield of low mass di-leptons\cite{CERES} have
been reported.
If these are due to thermal radiation from
hot matter,
an % Added due to referee comments A11
even larger production is
expected at RHIC energies, and can contribute to the % observed
electron signal.
Since $\rho \rightarrow e^+e^-$ contributes less than 1\% to the
calculated background as shown in Fig.~\ref{fig:ratio},
and since the dominant source of thermal di-leptons is
$\pi+\pi \rightarrow \rho \rightarrow e^{+}e^{-}$\cite{Rapp},
a significant contribution from thermal di-leptons is unlikely.
There are several predictions for
direct photons at RHIC energies\cite{Photon1,Photon2}.
The conversion electron spectrum calculated
from a prediction in ref.\cite{Photon1} is shown in
Fig.~\ref{fig:signal_spectrum} for central collisions.
% It can contribute
It could explain  % changed due to comment A12.
10-20\% of the signal, with large theoretical uncertainties.

%%%%%%%%%%%%%%%%%%%%%%%%%%%%%%%%%%%%%%%%%%%%%%%%%%%%%%%%%%%%%%%%
% P14 Estimate of the charm cross section
%
Neglecting these other possible sources and assuming that
all the electron signal is from charm,
we derive the charm cross section corresponding to the electron data.
We fit the charm electron spectrum from
PYTHIA to the % PHENIX 
data for $p_T>0.8$~GeV/c, and obtain the rapidity
density $dN_{c\bar{c}}/dy|_{y=0}$ and the total yield $N_{c\bar{c}}$
of open charm. They are then converted to cross sections
per $NN$ collision: 
$d\sigma_{c\bar{c}}/dy = (dN_{c\bar{c}}/dy)/T_{AA}$ and
$\sigma_{c\bar{c}} = N_{c\bar{c}}/T_{AA}$.
Results are shown % Ralf used ``listed''
in Table~\ref{tab:yield}.
%The systematic error is the
%quadratic sum of background subtraction (dominant) and
%dependencies on PYTHIA parameters, parton distribution functions,
%and fit range.
The systematic error is a quadratic sum of many sources. For central
collisions, they are background subtraction
($\pm 44\%$), uncertainties in the PYTHIA calculation
($\pm 11\%$ from $<k_T>=1.5 \pm 0.5$, $\pm 13\%$ from $D^+/D^0=0.65\pm0.35$,
$\pm 8\%$ from PDFs), fit range ($\pm 18\%$), and $T_{AA}$ ($\pm 7\%$).
Note that any finite contribution from neglected sources
would reduce the derived charm cross section.
Without nuclear or medium effects in charm production,
$\sigma_{c\bar{c}}$ per $NN$ collision should be independent of
centrality.
Within uncertainties, our data are consistent with
this expectation, in possible contrast to the attribution
of increased charm production as the source of enhanced di-muon
production reported in Pb+Pb collisions at SPS\cite{NA50b}.

\begin{figure}
%\begin{center}
\centerline{\epsfig{file=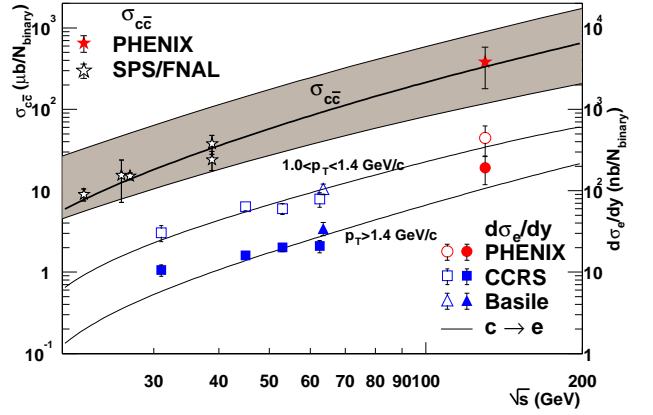,width=1.0\linewidth}}
\caption[]{
	Single electron cross sections $d\sigma_e/dy|_{y=0}$
	of this measurement and
	ISR experiments\cite{CCRS76,Basile,CCRS76b} are displayed
	(bottom of fig., right-hand scale)
	with charm decay contributions calculated with PYTHIA.
	Open and filled symbols
	are for $1.0<p_T<1.4$~GeV/c and $p_T>1.4$~GeV/c,
	respectively.
	The derived charm cross section of this measurement is compared
	with charm cross sections from SPS/FNAL experiments
	(top of fig., left-hand scale).
	The thick curve and the shaded band represent
	the charm cross section in the PYTHIA model 
	and in a NLO pQCD calculation\cite{MNR}, respectively.
	}
%\end{center}
\label{fig:comparison} 
\end{figure}

%%%%%%%%%%%%%%%%%%%%%%%%%%%%%%%%%%%%%%%%%%%%%%%%%%%%%%%%%%%%%%%%
% P15 Comparison with low energy data
%
The single electron signal yield (divided by
$T_{\small AA}$ to give the cross section per $NN$ collision)
and the derived charm cross section
are compared with single electron data of ISR experiments
and charm data of fixed target experiments\cite{charm_data}
in Fig.~\ref{fig:comparison}. 
Cross section curves calculated with PYTHIA, which has been
tuned to the charm data and the ISR electron data,
and a charm cross section curve from a
next-to-leading order (NLO)
pQCD calculation\cite{MNR} are also shown in the figure.
Our data are consistent with both of the calculations within large
uncertainties.

%%%%%%%%%%%%%%%%%%%%%%%%%%%%%%%%%%%%%%%%%%%%%%%%%%%%%%%%%%%%%%%%%
% P16 Summary and Outlook
%
In conclusion, we have observed single electrons above the expected
background from decays of light hadrons and photon
conversion in Au+Au collisions at $\sqrt{s_{_{NN}}}=130$~GeV.
The observed signal is consistent with semi-leptonic decay of charm.
The forthcoming high statistics Au+Au data and $pp$ comparison data
at full RHIC energy ($\sqrt{s_{_{NN}}}=200$~GeV)
will be useful to clarify the nature of the single electron signal and to
better determine heavy-quark production in Au+Au collisions at RHIC.

%%%%%%%%%%%%%%%%%%%%%%%%%%%%%%%%%%%%%%%%%%%%%%%%%%%%%%%%%%%%
% P16 Acknowledgments
%

%\section{Acknowledgements}

We thank the staff of the Collider-Accelerator and Physics Departments at
BNL for their vital contributions.  We acknowledge support from the
Department of Energy and NSF (U.S.A.), MEXT and JSPS (Japan), RAS,
RMAE, and RMS (Russia), BMBF, DAAD, and AvH (Germany), VR and KAW
(Sweden), MIST and NSERC (Canada), CNPq and FAPESP (Brazil), IN2P3/CNRS
(France), DAE and DST (India), KRF and CHEP (Korea), the U.S. CRDF for 
the FSU, and the US-Israel BSF.

%FIGURES:  Place all the figures here (after the references) in sequence.

%%%%%%%%%%%%%%%%%%%%%%%%%%%%%%%%%%%%%%%% Figure 1. (spectra)
\vspace{5cm}

%%%%%%%%%%%%%%%%%%%%%%%%%%%%%%%%%%%%%%%% Figure 2. (ratio)
\vspace{-0.5cm}

%%%%%%%%%%%%%%%%%%%%%%%%%%%%%%%%%%%%%%%% Figure 3. (xsection)
\vspace{-0.5cm}

%%%%%%%%%%%%%%%%%%%%%%%%%%%%%%%%%%%%%%%% Figure 4. (comparison)
\vspace{-0.5cm}

% TABLES
\begin{table}
\caption[]{ Charm cross section per $NN$ collision
            derived from the single electron data for central
            (0-10\%) and minimum bias (0-92\%) collisions.
	    The first and second errors are statistical and systematic,
	    respectively.
          }
\begin{tabular}[]{lccc}
Centrality& $T_{AA}$(mb$^{-1}$) & $d\sigma_{c\bar{c}}/dy|_{y=0}$($\mu$b)& $\sigma_{c\bar{c}}$($\mu$b) \\
\hline
0-10\% & $22.6\pm1.6$(sys.) & $97\pm13\pm49 $ & $380\pm60\pm200 $ \\
0-92\% &  $6.2\pm0.4$(sys.) & $107\pm8\pm63 $ & $420\pm33\pm250 $ \\
\end{tabular}
\label{tab:yield}
\end{table}

\end{multicols}


\begin{references}

\def\IJMPA{{Int. J. Mod. Phys.}~{\bf A}}
\def\JPG{{J. Phys}~{\bf G}}
\def\NCA{Nuovo Cimento}
\def\NIM{Nucl. Instrum. Methods}
\def\NIMA{{Nucl. Instrum. Methods}~{\bf A}}
\def\NPA{{Nucl. Phys.}~{\bf A}}
\def\NPB{{Nucl. Phys.}~{\bf B}}
\def\PL{{Phys. Lett.}~}
\def\PLB{{Phys. Lett.}~{\bf B}}
\def\PLC{Phys. Repts.\ }
\def\PRL{Phys. Rev. Lett.\ }
\def\PRD{{Phys. Rev.}~{\bf D}}
\def\PRC{{Phys. Rev.}~{\bf C}}
\def\ZPC{{Z. Phys.}~{\bf C}}
\def\EPJC{{Eur. Phys. J.}~{\bf C}}

\bibitem[*]{Deceased}Deceased     % for Leo Paffrath and Sten Garpman

\bibitem[\dagger]{non-par}Not a participating institution.

\bibitem{Appel} J.A.~Appel, Ann. Rev. Nucl. Part. Sci. {\bf 42}, 367 (1992).

\bibitem{XNWang} B.~M\"uller and X.N.~Wang, \PRL {\bf 68}, 2437 (1992).

\bibitem{Lin} Z.~Lin and M.~Gyulassy, \PRL {\bf 77}, 1222 (1996).

\bibitem{Kharzeev} Y.L.~Dokshitzer and D.E.~Kharzeev,
	\PLB {\bf 519}, 199 (2001).

\bibitem{Masui} T.~Matsui and H.~Satz,
	\PLB{\bf 178}, 416 (1986).

\bibitem{NA50} M.C.~Abreu {\it et al.}, \PLB {\bf 447}, 28 (2000).

\bibitem{Vogt} R.~Vogt {\it et al.},
	\PRD{\bf 49}, 3345 (1994).

\bibitem{CCRS74} F.~W.~B\"usser {\it et al.},
	\PL {\bf B53}, 212 (1974).

\bibitem{CCRS76} F.~W.~B\"usser {\it et al.},
	\NPB {\bf 113}, 189 (1976).

\bibitem{Perez} P.~Perez {\it et al.},
	\PL {\bf B112}, 260 (1982).

\bibitem{Basile} M.~Basile {\it et al.},
	Nuovo Cimento {\bf A65}, 421 (1981).

\bibitem{Hinchliffe_Smith} I.~Hinchliffe and C.~H.~Llewellyn Smith,
	\PL{\bf B61}, 472 (1976);
	M.~Bourquin and J.-M.~Gaillard, \NPB {\bf 114}, 334 (1976).

\bibitem{PPG003} K.~Adcox {\it et al.},
	\PRL {\bf 88}, 022301 (2002).

\bibitem{hamagakiQM01}  H.~Hamagaki {\it et al.},
                     \NPA{\bf 698}, 412 (2002).

\bibitem{PPG001} K.~Adcox {\it et al.},
	\PRL{\bf 86}, 3500 (2000).

\bibitem{GEANT} GEANT User's Guide, 3.15,
	CERN Program Library.

\bibitem{PPG006} K.~Adcox {\it et al.},
	nucl-ex/0112006.% submitted to \PRL

\bibitem{eta_ratio} R.~Albrecht {\it et al.},
	\PLB {\bf 361}, 14 (1995);
	G.~Agakichiev {\it et al.},
	\EPJC {\bf 4}, 249 (1998).% and references therein.	

\bibitem{omega_ratio} M.~Diakonou {\it et al.},
	\PL {\bf B89}, 432 (1980).
	
\bibitem{WA98} M.~M.~Aggarwal {\it et al.},
	nucl-ex/0006007;
	M.~M.~Aggarwal {\it et al.},
	\PRL {\bf 85}, 3595 (2000).

\bibitem{STAR_phi} C.~Adler {\it et al.},
	to be published.

\bibitem{PYTHIA} T.~Sjostrand, Comp. Phys. Commun.
	{\bf 82}, 74 (1994).

% \bibitem{CTEQ5} H.~L.~Lai {\it et al.}, \EPJC {\bf 12}, 375 (2000) 
\bibitem{PYTHIA_par}
We used PYTHIA 6.152 with CTEQ5L PDF
(H.~L.~Lai {\it et al.}, \EPJC {\bf 12}, 375 (2000)).
Modified PYTHIA parameters are: PARP(91)=1.5 ($<k_t>$), PMAS(4,1)=1.25 ($m_c$),
PARP(31)=3.5 ($K$~factor), MSTP(33)=1, MSTP(32)=4 ($Q^2$ scale).

\bibitem{charm_data} G.~A.~Alves {\it et al.},
	\PRL {\bf 77}, 2388 (1996).% and references therein.

\bibitem{CERES} G.~Agakichiev {\it et al.},
	\PLB {\bf 422}, 405 (1998).

\bibitem{Rapp} R. Rapp, \PRC {\bf 63}, 054907 (2001).

\bibitem{Photon1} J. Alam {\it et al.}, \PRC {\bf 63}, 021901 (2001).

\bibitem{Photon2}
	F.~D.~Steffen and M.~H.~Thoma, \PLB {\bf 510}, 98 (2001);
	D.~K.~Srivastava, nucl-th/0103023.

\bibitem{NA50b} M.C.~Abreu {\it et al.}, \EPJC {\bf 14}, 443 (2000).

\bibitem{MNR} M.~Mangano, P.~Nason, and G.~Ridolfi,
	\NPB {\bf 405}, 507 (1993). Their % Fortran
        program {\bf HVQMNR} is used with CTEQ5M
	% parton distribution functions
	PDF
	to calculate $\sigma_{c\bar{c}}$ in
	Fig.~\ref{fig:comparison}
	with $m_c=1.5$ GeV/$c^2$, $\mu_F = 2m_c$, and
	$0.5m_c < \mu_R < 2m_c$.

\bibitem{CCRS76b} The 1.0-1.4 GeV/c point of CCRS is calculated from
	the $e/\pi$ ratio in ref.\cite{CCRS76} and the pion cross section in
	B.~Alper {\it et al.}, \NPB {\bf 100}, 237 (1975).

\end{references}
\end{document}